\documentclass[12pt]{iopart}

\usepackage{iopams}  
\usepackage{tabularx,ltxtable,graphicx,epsfig,harvard}
\begin{document}

\note[Heart rate scaling graphs in athletes]{Scaling graphs of heart rate time series in athletes demonstrate the VLF, LF and HF regions}

\author{Mathias~Baumert$^{1,3,4}$,  Lars~M~Brechtel$^2$, Juergen~Lock$^2$, Andreas~Voss$^3$ and Derek~Abbott$^{1,4}$}

\address{Centre for Biomedical Engineering (CBME), The University of Adelaide, SA 5005, Australia}
\address{Department of Sports Medicine, Humboldt-University Berlin, Fritz-Lesch-Strasse 29, 13053 Berlin, Germany}
\address{Department of Medical Engineering, University of Applied Sciences Jena, Carl-Zeiss-Promenade 2, 07745 Jena, Germany}
\address{School of Electrical \& Electronic Engineering, The University of Adelaide, SA 5005, Australia}
\ead{mbaumert@eleceng.adelaide.edu.au}

\begin{abstract}
Scaling analysis of heart rate time series has emerged as an useful tool for assessment of autonomic cardiac control. We investigate the heart rate time series of ten athletes (five males and five females), by applying detrended fluctuation analysis (DFA). High resolution ECGs are recorded under standardized resting conditions over 30 minutes and subsequently heart rate time series are extracted and artefacts filtered. We find three distinct regions of scale-invariance, which correspond to the well-known VLF, LF, and HF bands in the power spectra of heart rate variability. The scaling exponents $\alpha$ are $\alpha_{\rm{HF}}$: 1.15 [0.96--1.22], $\alpha_{\rm{LF}}$: 0.68 [0.57--0.84],  $\alpha_{\rm{VLF}}$: 0.83[0.82--0.99]; $p<10^{-5}$). In conclusion, DFA scaling exponents of heart rate time series should be fitted to the VLF, LF, and HF ranges, respectively.
\end{abstract}


\maketitle

\section{Introduction}
Scaling analysis of heart rate time series has emerged as an useful tool for assessment of autonomic cardiac control and has been shown to be useful for diagnostics in patients with cardiac disease \cite{Ho:97}. One widely applied approach, for the investigation of scaling characteristics, is detrended fluctuation analysis (DFA) \cite{Peng:95}. Scale-invariance has been commonly observed over a wide range with a characteristic break at segment sizes of 16 heart beats. Consequently, two scaling exponents, termed $\alpha_1$ and $\alpha_2$, are computed in the ranges of 4--16 and 16--64 heart beats, respectively. Early investigators considered fractal scaling analysis as a means to providing a unique view into autonomic control \cite{Peng:95}. Recently, however, the mathematical link between DFA and the classical power spectrum analysis of heart rate variability (HRV) has been provided \cite{Willson:02}, \cite{Willson:03} and a physiological interpretation of $\alpha_1$ and $\alpha_2$ in the framework of the well-studied VLF, LF and HF frequency bands has been given \cite{Francis:02}. Nevertheless, the fractal nature of HRV remains an open question, especially when the state of the autonomous nervous system is altered by exercise \cite{Karasik:02}.

\section{Methods}
Roughly speaking, DFA relates the variance in a detrended time series versus the size of the linearly trend eliminated segments. Note that DFA has been developed to analyze long-range correlations (long-memory dependence) in non-stationary data, where conventional fluctuation analyses such as power spectra and Hurst analysis cannot be reliably used \cite{Peng:95}. The method works as follows:
\begin{itemize}
\item	Compute the cumulative sum $c(k)=\sum_{i=1}^k[s(i)-\overline{s}]$ of the zero-mean (beat-to-beat RR interval) time series, where $\overline{s}$ denotes the mean of the time series $s$ (using the concept of Random-Walk-Analysis \cite{Peng:95}).
\item	Compute the local trend $c_n(k)$ within boxes of varying sizes $n$ (linear least square fit).
\item	Compute the root-mean-square of the detrended time series in dependency on box size $n$ as  $F(n)=\sqrt{\frac{1}{N}\sum_{k=1}^N[c(k)-c_n(k)]^2}$, where $N$ denotes the length of $s$.
\item Plot $\log_{10} F(n)$ against $\log_{10} n$.
\end{itemize}
If the data displays long-range dependence, then $F(n) \propto n^\alpha$---where $\alpha$ is the scaling exponent that is obtained via least square fit. For stationary data with scale-invariant temporal organization, the Fourier power spectrum $S(f)$ is $S(f) \propto f^{-\beta}$, where the scaling exponent $\beta$ is related to a in the following way: $\beta = 2\alpha -1$ \cite{Peng:93,Peng:94}. Thus time series with $1/f$ in the power spectrum (i.e.~$\beta=1$) are characterized by DFA exponent $\alpha=1$. Values of $0 < \alpha < 0.5$ are associated with anti-correlation (i.e.~large and small values of the time series are likely to alternate). For Gaussian white noise $\alpha = 0.5$. Values of $0.5 < \alpha \leq 1$ indicate long-range power-law correlations (i.e. large values of the time series are likely to be followed by large values) and describe the decay $\gamma$ of auto-correlation function $C(n)\equiv\left\langle s(i)s(i+n)\right\rangle \propto n^{-\gamma}$, where $\gamma=2-2\alpha$. Values $1 < \alpha \leq 1.5$ represent stronger long-range correlations that are different from power-law, where $\alpha = 1.5$ for Brownian motion \cite{Peng:95}.
\\

We perform DFA in heart rate time series of ten healthy experienced athletes (five males and five females) from track and field as well as triathlon. Anthropometric data and peak oxygen uptake are shown in table~\ref{Tab:1}, applying non-parametric statistics; all subjects being fully recovered from the competition season and the results of a medical examination were negative. No athlete received medication prior to the study. High resolution ECGs (1600 Hertz) are recorded in a supine position under standardized resting conditions over 30 minutes. Heart rate time series are automatically extracted and artifacts are subsequently filtered and interpolated based on local variance estimation. The investigation conforms to the principles outlined in the Declaration of Helsinki, with written informed consent of all athletes being provided.
\begin{table}
\caption{\label{Tab:1}Anthropometric data and peak oxygen uptake of the ten investigated athletes presented as medians and interquartile ranges (IQR).}
\begin{indented}
\item[]\begin{tabular}{@{}lllll}
\br
Sex	&\multicolumn{2}{c}{Women}	&\multicolumn{2}{c}{Men}\\ 
&Median	&IQR	&Median	&IQR\\
\mr
Age [years]	&24.8	&24.7--26.4	&26.6	&26.5--28.8\\
Body mass [kg]	&54.8	&50.4--61.8	&72.0	&69.0--86.8\\
Height [cm]	&163	&162--168	&181	&181--182\\
Body fat [\%]	&18.0	&15.0--24.0	&14.0	&12.0--21.0\\
VO$_{2}$ peak [ml*(kg*min)$^{-1}$]	&51.1	&48.9--52.2	&65.9	&61.4--74.6\\
\br
\end{tabular}
\end{indented}
\end{table}
\section{Results and Discussion}
\begin{figure}%[htbp]
\centering 
\epsfig{file=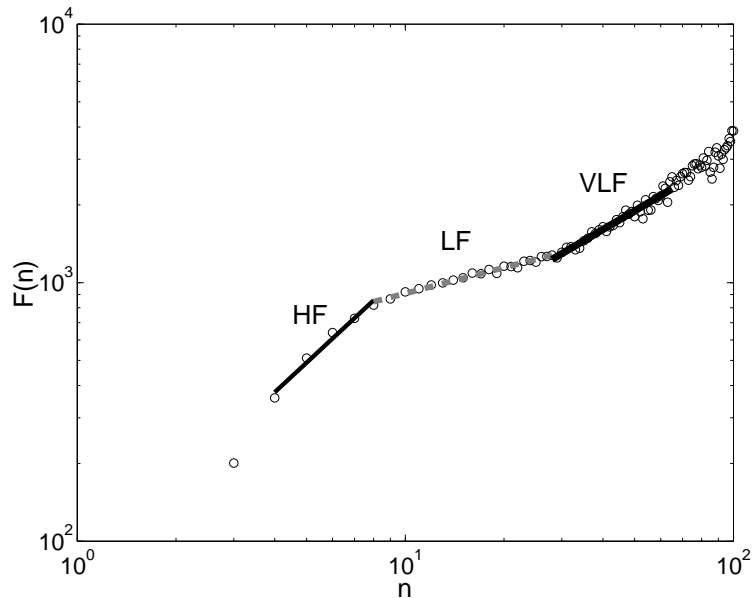, width=10 cm}
%\includegraphics[hight=3cm]{wortverteilung.png}
%\centering{\frame{\rule{2cm}{2cm}}}
\caption{Detrended fluctuation analysis of heart rate time series is performed over a sample of ten athletes. This figure represents one example plot, for one athlete, clearly demonstrating for the first time the three distinct frequency regions. $F(n)$---root-mean-square of the detrended time series. Here, $n$---segment size for linear trend elimination. Solid line---scaling range equivalent to the HF band, starting with $n=4$; dotted gray line---scaling range equivalent to the LF band; bold line---scaling range equivalent to the VLF band, truncated at $n=64$.} 
\label{Fig:1}
\end{figure}

In most subjects, DFA reveals three distinct regions of scale-invariance as seen in figure~\ref{Fig:1}. By investigating the location of the breakpoints, we find a direct relationship between the well-known HRV frequency bands (a) very low frequency (VLF): 0.--0.04 Hertz, (b) low frequency (LF): 0.04--0.15 Hertz, and (c) high frequency (HF): 0.15--0.4 Hertz \cite{HRV-TF}. The two break points between the three regions of scale-invariance correspond to the border frequencies of power spectrum analysis, i.e. 0.04 and 0.15 Hertz. In order to relate frequency values $f_n$ in Hertz from the segment size $n$ of DFA, we use the rough approximation: $f_n\approx(\overline{s}n)^{-1}$. Thus, the three scaling regions are individually computed, depending on the traditional two fixed border frequencies and the individual mean heart rate. Lower and upper boundaries for the analysis are $n = 4$ and $n = 64$, respectively, as proposed in the original work by Peng {\it et al.} \cite{Peng:95}. %The corresponding frequency value $f_n$ in the Discrete Fourier Transform power spectrum is then $f_n=n(N \Delta t)^{-1}$. 
\\

\begin{figure}%[htbp]
\centering 
\epsfig{file=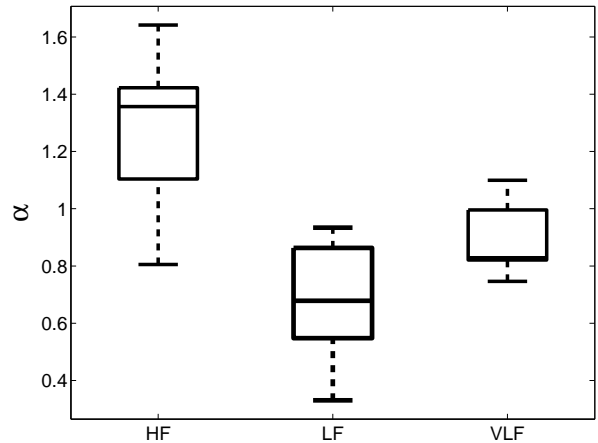, width=8 cm}
%\includegraphics[hight=3cm]{wortverteilung.png}
%\centering{\frame{\rule{2cm}{2cm}}}
\caption{Boxplots of the HF-, LF-, and VLF- range scaling exponents obtained from heart rate time series of all ten athletes by applying detrended fluctuation analysis. The Friedman test indicates highly significant group median differences ($p < 10^{-5}$). Values displayed by the box plot are (a) the median, indicated inside each box, (b) the interquartile range, indicated as the height of each box, and (c) the whiskers indicate the total range of the data.} 
\label{Fig:2}
\end{figure}

All three scaling exponents are significantly different from each other (see figure~\ref{Fig:2}, Friedman test for non-parametric group median comparisons of repeated measurements: $p < 10^{-5}$). The HF scaling exponents of all subjects indicate the presence of correlation, i.e.~$\alpha > 0.5$, and might be caused by the strictly periodic nature of respiration. The LF scaling exponent---usually associated with sympathetic, vagal cardiac and vascular control---reveals a much less strict long-term correlation and therefore a less strict control regime. In individual cases, the correlation disappears, i.e.~$\alpha = 0.5$, or even becomes anti-correlated, i.e. $\alpha < 0.5$. The VLF exponent shows long-term correlation, in all subjects, but lower than those of the HF range. Although a physiological explanation of VLF oscillations is still under debate, links with the renin-angiotensin-system, for example, have been suggested \cite{Bonaduce:94}. 
Our analysis clearly shows that DFA scale-invariance of HRV is directly linked to well-known physiological phenomena of VLF, LF, and HF oscillations. Although this relationship has been shown mathematically before \cite{Willson:02}, \cite{Willson:03}, we are for the first time able to show this relationship directly in scaling graphs, such as the example given in figure~\ref{Fig:1}. The initial scale invariance ($\alpha_1$), which has been reported often \cite{Ho:97}, \cite{Peng:95}, \cite{Karasik:02}, seems therefore to be predominately caused by respiratory modulations---whereas the $\alpha_2$ range of scale invariance reflects the typical $1/f$ characteristics of VLF oscillations, which have been shown in HRV power spectra \cite{HRV-TF}. The ´crossover range' between those areas of scale invariance \cite{Karasik:02} is caused by LF oscillations that show only weak correlation. Interestingly, three scaling exponents were found in the beat-to-beat blood pressure dynamics of normal inactive subjects as well as in patients with dilated cardiomyopathy \cite{Baumert:05}. Obviously, the relatively low mean heart rate in athletes (58 [50-64] beats per minute) reveals the separate VLF, LF and HF scaling regions, while these might remain partly masked in untrained probands. Further, a change in cardiac activity mediated by sympathetic and vagal efferents, as has been observed in athletes \cite{Carter:03}, might play a role. Therefore, scaling exponents and graphs, respectively, could provide an additional tool in monitoring the effects of physical activity on the regulation of the autonomous nervous system in patients and healthy subjects and in the monitoring of athlete training.
\\´

We conclude that DFA scaling exponents of HRV should be fitted to three ranges; namely the VLF, LF, and HF ranges, respectively.

\ack
This study was partly supported by grants from the Deutsche Forschungsgemeinschaft (DFG Vo505/4-2), from the Australian Research Council (DP0663345), from the Stiftung Warentest Berlin, and from the SMS---Sports Medicine Service, Berlin, Germany.

\section*{References}
\providecommand{\SortNoop}[1]{}

\end{document}